\documentclass[preprint, 3p, review, authoryear]{elsarticle}

\usepackage{amsmath}
\usepackage{amssymb}
\usepackage{amsfonts}
\usepackage{amsthm}
\usepackage[T1]{fontenc}
\usepackage[utf8]{inputenc}
\usepackage{graphicx}
\usepackage{float}
\usepackage{caption}
\usepackage{subcaption}
\usepackage{booktabs}
\usepackage{siunitx}
\usepackage{url}

\def\removeSubmitedTo{1}

\if\removeSubmitedTo1
\makeatletter
\def\ps@pprintTitle{%
  \let\@oddhead\@empty
  \let\@evenhead\@empty
  \let\@oddfoot\@empty
  \let\@evenfoot\@oddfoot
}
\makeatother
\fi

\numberwithin{equation}{section}

\newtheorem{thm}{Theorem}
\newtheorem{coro}{Corollary}
\newtheorem{lemma}{Lemma}
\newproof{pf}{Proof}

\begin{document}
\begin{frontmatter}



\title{Brazilian listed options with discrete dividends and the fast Laplace transform}


\author{Maikon Araújo}

\affiliation{organization={IME-USP},
            addressline={Rua do Matão, 1010}, 
            city={São Paulo},
            postcode={05508-090}, 
            state={São Paulo},
            country={Brazil}}

\begin{abstract}
The Brazilian stock exchange (B3) has long used a strike-only adjustment to account for 
dividends in its listed equity options. 
This adjustment still makes it necessary to account for discrete dividends
when pricing either calls or puts. This work presents a numerical procedure,
based on the fast Laplace transform and its inverse, a procedure that can efficiently 
compute the Brazilian listed options' premium and the Greeks delta, gamma, 
and theta with high accuracy.
\end{abstract}


\begin{highlights}
\item Pricing method for European options with discrete dividends.
\item Numerical method using the fast Laplace transform.
\item Fast Laplace transform via fast Fourier transform.
\item Option's sensitivities.
\end{highlights}

\begin{abstract}
The Brazilian stock exchange (B3) has long used a strike-only adjustment to account for 
dividends in its listed equity options. 
This adjustment still makes it necessary to account for discrete dividends
when pricing either calls or puts. This work presents a numerical procedure,
based on the fast Laplace transform and its inverse, a procedure that can efficiently 
compute the Brazilian listed options' premium and the Greeks delta, gamma, 
and theta with high accuracy.
\end{abstract}    

\end{frontmatter}


\section{Introduction}

In the Brazilian market, on each ex-dividend date, listed equity options have their strike reduced 
by the respective dividend value.
Merton has proved in his paper \cite[p. 152]{merton1973rationaloptionpricing}
that options under this adjustment are not dividend protected, therefore it is necessary
to incorporate these discrete dividends to price these options. 

To account for discrete dividends, many authors had made great contributions in the realm of
approximations and numerical solutions. In the book \cite{haug2007complete},
Haug compares many approximations and adjustments from other authors and gives an excellent 
alternative to the case of multiple dividends also presented in \cite{haug2003back}.
In an outstanding work, \cite{THAKOOR20181} advocate for the use of Fejér quadrature to 
approach the integration step in each dividend payment date, reporting that such implementation 
yields fast and accurate results.

This work presents a new alternative to solve the multiple dividends problem in 
the case of listed Brazilian options. 
Section \ref{sec:dynamics} lays the theoretical foundation,
which consists of one way to model the impacts of dividends on the price of Brazilian 
call and put options. Brazilian listed options are constituted of calls with American and European 
exercises and puts with European exercise, this section also proves that both options can be treated as having European 
exercise for pricing purposes, and gives a pricing strategy 
based on the Laplace transform and its inverse.
Section \ref{sec:numerical_procedure} delivers the main result of this paper,
formulating a numerical procedure to price listed Brazilian options, explaining how to implement 
the fast Laplace transform (FLT) as a function that calls the already known fast Fourier transform, 
and driving through the discrete approximation details to find the option's premium and sensitivities 
to the underlying and to time. Section \ref{sec:results} shows that the FLT pricer delivers 
the option's premium and Greeks with high accuracy and performance when compared with the
benchmark implementation from \cite{THAKOOR20181}.

The idea of building a fast Laplace transform via the fast Fourier transform 
presented in section \ref{sec:numerical_procedure} was inspired 
by a great online video from \cite{yt_steven_brunton_laplace} where he
explains how to derive the Laplace transform as an extension of the Fourier transform.
More on the subject of using the spectral derivatives to solve the heat equation can be found in 
the book \cite{brunton2022data}.

\section{Option dynamics}
\label{sec:dynamics}

The dividend dynamic has the impact of lowering the stock price by the present value of the dividend at the ex-dividend date. 
If the owner sells a stock before the ex-dividend date the future dividend payment goes to the stock buyer,
on the other hand, if the stock is sold after this date the dividend will go instead 
to the person who owned the stock at the ex-dividend date, even if the payment is set to occur on a later date. 
Hence, the new buyer will expect to pay a lower price because the stock owner just lost the right to receive that 
dividend payment.

For options, this drop in value has the effect of reducing the value of call options and raising the value of put options.
This is the result of a direct impact on the intrinsic value of these options, namely:
\begin{align*}
    \text{call} = \max\{S_t - K, 0\},\;     \text{and put} = \max\{K - S_t , 0\}.
\end{align*}

To protect option holders from this price change, the Brazilian stock exchange adjusts the option's strike 
price by subtracting the dividend value from the strike on the same ex-dividend date, which keeps the intrinsic value for both 
option types. As Merton has shown in his paper \cite[p. 152]{merton1973rationaloptionpricing}, although the 
intrinsic value is preserved the option price still depends on the value and ex-date of the dividend.

This section will cover the theoretical background and main theorems used to model the option's dynamics
and price over time.

\subsection{Theoretical background}

Let $f(t)$ be right continuous with left limits
such that $f(t_-) = \lim_{u \uparrow t}f(u)$,
and $\mathcal J f_t = f(t) - f(t_-)$. Note that if $f$ is
continuous in the open interval $(a,b)$ thus $\mathcal J f = 0$.
With these definitions, the stochastic differential equation (SDE)
for a dividend-paying stock can be expressed as:

\begin{subequations}
    \label{eq:s_process_with_dividend}
    \begin{align}
        dS_t &= S_t r_t \mathrm dt + S_t\sigma_t \mathrm dW_t + \mathcal J  S_t, \\
        \mathcal J S_t &=
        \left\lbrace \begin{array}{rl}
            -D_i, & \text{if } t = t_i \text{, and} \\
            0,    & \text{otherwise.}
        \end{array}\right.
    \end{align}
\end{subequations}

Where $ 0 < t_1 < \dots < t_n < T$, with $\{t_1, \dots, t_n\}$ being ex-dividend dates,
and $\{D_1, \dots, D_n\}$ being the value of each dividend paid respectively. To simplify
the notation in this work is assumed that the payment date occurs on the same ex-date, this 
assumption can be weakened  by taking $D_i$ to be the discounted value from all parts of that dividend 
payed on dates $t^\star_l$ with amounts of $D^\star_l$  to the ex-date $t_i$:
\begin{align} \label{eq:div_present_value}
    D_i = \sum_l  D^\star_l \exp\left(-\int_{t_i}^{t^\star_l}r_u\; \mathrm du\right).
\end{align}

It is also useful to define a money account where the dividend paid at each time $t_i$ is invested.
The value of this account at any time can be expressed as:
\begin{subequations}
    \label{eq:cash_past_dividends}
    \begin{align}
        I_t &= \sum_{i=1}^n D_i  \exp \left( {\int_{t_i}^t r_u\; \mathrm du}\right) 1_{[t_t, \infty)}(t), \\
        \mathrm dI_t &= r_t I_t \mathrm  dt + \mathcal  J I_t, \\
        \mathcal J I_t &=   \left\lbrace \begin{array}{rl}
            D_i, & \text{if } t = t_i \text{, and} \\
            0,   & \text{otherwise.}
        \end{array}\right.
    \end{align}
\end{subequations}

\begin{thm} \label{thm:black_scholes_pde}
    Between ex-dividend dates, $t \in [t_k, t_{k+1})$, a security $V_t = V(S_t, t)$ possesses the following
    partial differential equation (PDE):
    \begin{align}
        \label{eq:black_scholes_pde}
        \frac{\partial V_t}{\partial t}
        + \frac{(\sigma_t S_t)^2}2  \frac{\partial^2 V_t}{\partial S_t^2}
        +  S_t r_t \frac{\partial V_t}{\partial S_t} -  r_tV_t  = 0,
    \end{align}
    which is the standard PDE from \cite[p. 643, eq. 7]{scholes1973pricing} for a
    non-dividend-paying stock.
\end{thm}

The derivatives in equation \eqref{eq:black_scholes_pde} are also referred as Greeks 
, namely $\Delta$, $\Gamma$, and $\Theta$, such that:
\begin{align}
    \Delta = \frac{\partial V_t}{\partial S_t}, \; 
    \Gamma=\frac{\partial^2 V_t}{\partial S_t^2},\; \text{ and }
    \Theta = \frac{\partial V_t}{\partial t}.
\end{align}

\begin{coro} \label{coro:option_martingale}
    From theorem \ref{thm:black_scholes_pde}, the discounted security price
    $X_t = e^{-\int_0^t r_u\;\mathrm du}V_t$ is a martingale.
\end{coro}

\begin{coro} \label{coro:put_price}
    From corollary \ref{coro:option_martingale}, the price of a European put option with strike adjustment is:
    \begin{align}
        P_t = \mathbb E \left[ \left.  e^{-\int_t^T r_u\;\mathrm  du} \max\left\{K_T - S_T, 0\right\}\right| \mathcal F_t\right].
    \end{align}    
\end{coro}

\begin{coro}
    If $r$ and $\sigma$ are constants and with some constant $K_T$,
    usually the strike price for vanilla options,
    the following change of variables:
    \begin{subequations}
        \label{eq:pde_change_of_variables}
        \begin{align}
            \tau &= T - t,\;
            \tilde S_\tau = S_t,  \label{eq:pde_change_tau_S}\\
            v(x, \tau )   &= \tilde V(\tilde S_\tau, \tau) = V(S_t, t) \label{eq:pde_change_Vv}\\
            x &= \ln \frac {\tilde S_\tau}{K_T} + \left(r - \frac{\sigma^2}{2}\right) \tau, \label{eq:pde_change_Sx}\\
            F(x ,\tau) &= v(x, \tau)e^{\int_0^\tau r_u\;\mathrm du}, \label{eq:pde_change_vF}
        \end{align}
    \end{subequations}
    imply the derivatives:
    \begin{subequations}
        \label{eq:black_heat_equation_greeks}
        \begin{align} 
            \frac{\partial V_t}{\partial S_t} &= \frac 1 S_t \frac{\partial F}{\partial x} e^{-\int_t^T r_u\;\mathrm  du},\label{eq:black_heat_equation_greeks:delta}\\
            \frac{\partial^2 V_t}{\partial S_t^2} &= \frac{1}{S_t^2}  \left( \frac{\partial^2 F}{\partial x^2} - \frac{\partial F}{\partial x}\right)e^{-\int_t^T r_u\;\mathrm  du},\label{eq:black_heat_equation_greeks:gamma}\\
            \frac{\partial V_t}{\partial t} &=  - \left( \frac{\partial F}{\partial\tau} +\left(r - \frac{\sigma^2}{2}\right) \frac{\partial F}{\partial x}   - r_t F  \right)e^{-\int_t^T r_u\;\mathrm  du},
        \end{align}
    \end{subequations}
    which transform equation \eqref{eq:black_scholes_pde} into a standard heat equation:
    \begin{align} \label{eq:black_heat_equation}
        \frac{\partial F}{\partial \tau} = \frac{\sigma^2} 2 \frac{\partial^2 F}{\partial x^2}.
    \end{align}
\end{coro}

Theorem \ref{thm:black_scholes_pde} allows derivatives on dividend-paying stocks to be treated the same way
derivatives on a stock without dividends, when $t$ is between ex-dividends dates.
The only distinction is at the exact moment $t = t_i$ when a dividend goes ex date $(t_i^-\rightarrow t_i)$.
Here is assumed that, although the stock price has jumped, the security price remains the same. To
represent this assumption the price $V_t$ is remapped into the new stock price so that:
\begin{align} \label{eq:security_dividend_remapping}
    V_{i-1}(S_{t_i^-},{t_i^-}) = V_i(S_t, t_i),
\end{align}
where $V_i(S_t, t)$ is the security price for $t \in [t_i, t_{i+1})$.

In particular, for American call options, the strike adjustment on ex-dates works in a similar fashion
preserving the intrinsic value at jumps. To see that let $g_i(S_t) = \max(S_t - K_{t_i}, 0)$ be the 
intrinsic value of a call option at time $t \in [t_i, t_{i+1})$:
\begin{align}
    g_{i-1}(S_{t_i^-}) &= \max(S_{t_i^-} - K_{t_i^-}, 0)  \nonumber \\
    &= \max(S_{t_i^-} -D_i - (K_{t_i^-} - D_i), 0)  \nonumber \\
    &= \max(S_{t_i} - K_{t_i} , 0),  \nonumber \\
    \therefore \, g_{i-1}(S_{t_i^-})&= g_i(S_{t_i^-}), \label{eq:call_payof_continuous}
\end{align}
with  $K_0 = K$ being the initial strike price, and:
\begin{align}
    K_{t_i} = K_{t_i^-} - D_i,
\end{align}
being the adjustment made on this strike at the ex-date. 


The fact that the intrinsic value is preserved makes the American call option behaves like
a European option as the intrinsic value is always less than the security price.
Therefore:
\begin{thm} \label{thm:american_is_european}
    The early exercise of Brazilian listed call options is never optimal
    \footnote{The option owner can always exercise early if it is desired, but it is never optimal.}
    because these options are strike adjusted.
    So American calls can be treated as having European exercise for pricing purposes.
\end{thm}

\begin{coro} \label{coro:call_price}
    From theorem \ref{thm:american_is_european} and corollary \ref{coro:option_martingale}, the 
    price of a call option with strike adjustment is:
    \begin{align}
        C_t = \mathbb E \left[ \left.  e^{-\int_t^T r_u \;\mathrm du} \max\left\{S_T - K_T, 0\right\}\right| \mathcal F_t\right].
    \end{align}    
\end{coro}

\begin{coro}
    \label{coro:parity_put_call}
    From theorem corollaries \ref{coro:call_price} and \ref{coro:put_price} follows that the put-call parity holds
    for dividend adjusted options:
    \begin{align}
        \label{eq:parity_put_call}
        C_t - P_t = \mathbb E \left[ \left.  e^{-\int_t^T r_u\;\mathrm  du} \left(S_T - K_T\right)\right| \mathcal F_t\right].
    \end{align}
\end{coro}

\begin{thm}
    \label{thm:forward_price}
    If the interest rate $r_t$, dividend dates $t_i$ and values $D_i$, are deterministic,
    the forward price is given by:
    \begin{align}\label{eq:forward_price}
        \mathbb E\left[\left. S_T  \right| \mathcal F_t \right] = e^{\int_t^T r_u\;\mathrm  du}(S_t + I_t) - I_T.
    \end{align}
\end{thm}

\subsection{Pricing strategy}
\label{sec:pricing_strategy}

To price European vanilla options
on dividend paying stocks, equation \eqref{eq:black_heat_equation} is solved recursively in each interval
from  $0 < \tau_1 < \dots < \tau_n < T $, where $\tau_i = T - t_{n - i + 1}$,
and $\tau_i^- = T - t_{n 1 - i }^-$. To do so,
consider a distinct function $F_i(x, \tau)$ for each interval $(\tau_{i-1}^-, \tau_i]$,
adding one more assumption that the dividends $D_i$ are known at time $t=0$ and solve
${n+1}$ PDEs such that:
\begin{align}    \label{eq:pde_recursive}
    \frac{\partial F_i}{\partial \tau} = \frac{\sigma^2} 2 \frac{\partial^2 F_i}{\partial x^2},
\end{align}
with the following initial conditions:
\begin{align}
    F_i(x_{\tau_{i-1}^-}, \tau_{i-1}^-) &= F_{i-1}(x_{\tau_{i-1}}, \tau_{i-1}),
\end{align}
and especially for call options, the initial and boundary conditions are:
\begin{subequations}
    \label{eq:initial_and_boundary_call_conditions}
    \begin{align}
        F_1(x_{0}, 0) &= K_T\max\left\{e^{x_0} - 1, 0\right\} ,  \label{eq:call_ic}\\
        \left. \frac {\partial}{\partial x}F_i(x, \tau)  \right|_{x=0} &= F_i(0, \tau) = 0. \label{eq:call_bc}
    \end{align}
\end{subequations}
Solving for the call option price the put option price can be found using the put-call parity and the forward price from
equations \eqref{eq:parity_put_call} and \eqref{eq:forward_price}.

Now, take the Laplace transform for $x$ on both sides of the equation \eqref{eq:pde_recursive}:
\begin{align}
    \mathcal L\left\{ \frac{\partial F_i(x,\tau)}{\partial \tau} \right\} &= \mathcal L\left\{ \frac{\sigma^2} 2 \frac{\partial^2 F_i(x, \tau)}{\partial x^2} \right\} , \nonumber \\
    \frac{\partial \bar F_i(s, \tau)}{\partial \tau}  &=  \frac{\sigma^2} 2 \frac{\partial^2 \bar F_i(s, \tau)}{\partial x^2} \nonumber \\
    &=  \frac{\sigma^2} 2 \left[
    s^2\bar F_i(s, \tau) -
    \left(
    s F_i(x, \tau) +
    \left. \frac {\partial}{\partial x}F_i(x, \tau)
    \right)\right|_{x=0^+}
    \right], \label{eq:laplace_diff_heat_equation}
\end{align}
applying the boundary conditions \eqref{eq:call_bc}:
\begin{align}
    \frac{\partial \bar F_i(s, \tau)}{\partial \tau}
    &=  \frac{(\sigma s)^2} 2
    \bar F_i(s, \tau),
\end{align}
which can be solved for $\tau_i$ starting from $\tau_{i-1}^-$:
\begin{align} \label{eq:pde_s_solution}
    \bar F_i(s, \tau_i)
    &=  e^{\frac{(\sigma s)^2} 2  (\tau_i - \tau_{i-1})}
    \bar F_i(s, \tau_{i-1}^-),
\end{align}
invert the Laplace transform to get to $x$ domain:
\begin{align} \label{eq:pde_x_solution}
    F_i(x, \tau_i)
    &=  \mathcal L^{-1}\left\{  \bar F_i(s, \tau_i) \right\},
\end{align}
and then remapped to account for the dividend event:
\begin{align} \label{eq:f_dividend_remapping}
    F_{i+1}(x_{\tau_i^-},{\tau_i^-}) = F_{i}(x_\tau, \tau_i).
\end{align}
Repeat the steps from equation \eqref{eq:pde_s_solution} to \eqref{eq:f_dividend_remapping}
until  $F_{n+1}(x_T, T)$ is found. Finally, use equations \eqref{eq:pde_change_tau_S}  to \eqref{eq:pde_change_vF}
to retrieve $V_{n+1}(S_0, 0)$ which is the call option price. The Greeks $\Delta$ and $\Gamma$
are found by replacing:
\begin{subequations}
    \label{eq:fx_derivatives}    
    \begin{align}
        \frac{\partial F_{n+1}(x_T, T)} {\partial x} &= \mathcal L^{-1}\left\{ s \bar F_{n+1}(s, T) \right\}, \\
        \frac{\partial^2 F_{n+1}(x_T, T)} {\partial x^2} &= \mathcal L^{-1}\left\{ s^2 \bar F_{n+1}(s, T) \right\}
    \end{align}
\end{subequations}
in equations \eqref{eq:black_heat_equation_greeks:delta} and \eqref{eq:black_heat_equation_greeks:gamma}
respectively, and $\Theta$ can be found by replacing $\Delta$ and $\Gamma$ into the equation 
\eqref{eq:black_scholes_pde}.

\section{Numerical procedure}
\label{sec:numerical_procedure}

This section brings the main result of this paper, a pricing procedure that can give the theoretical price for Brazilian listed
equity options in the presence of dividends. Keeping the same assumptions from \cite{scholes1973pricing}.
The idea consists of reducing the PDE in equation \eqref{eq:black_scholes_pde} to the heat
equation \eqref{eq:pde_recursive} that can be solved numerically with the fast Laplace transform, taking special care for each
ex-date where we remap the spacial variable as described in equation \eqref{eq:security_dividend_remapping}.

It is customary to reduce equation \eqref{eq:pde_recursive} to an ordinary differential equation by applying the Fourier
transform in the $x$ domain to solve the heat equation analytically, unfortunately,
this will make us run into numerical problems for securities like vanilla calls
because their price function is not periodic, therefore, not well suitable for a Fourier transform (FFT).
To solve this problem the fast Laplace transform is used, a way to leverage the fast Fourier transform
implementation to work on problems where the Laplace transform is better suited.

\subsection{Fast Laplace transform}

To understand how the Fourier transform can be used to derive the Laplace transform, imagine a
function $f(x)$ that is not periodic. To use the Fourier transform
work on a new function $g(x;\lambda)$, such that:
\begin{align} \label{eq:flt_gx}
    g(x; \lambda) = e^{-\lambda x} f(x) u(x),
\end{align}
where $u(x)$ is the Heaviside step function, and $\lambda$ is some positive value that will
make $e^{-\lambda x} f(x) \to f(0)$ when $x \to \infty$, so that $g(x; \lambda)$ can be considered periodic on 
that interval. Applying the Fourier transform to both sides of equation \eqref{eq:flt_gx} gives:
\begin{align}
    \mathcal F\left\{ g(x) \right\} &= \int_{-\infty}^\infty e^{-i 2\pi \xi  x} e^{-\lambda x} f(x) u(x)\;\mathrm  dx, \\
    \hat g(\xi; \lambda) &= \int_0^\infty e^{-\left(\lambda + i2\pi\xi \right) x}  f(x) \;\mathrm  dx,
\end{align}
which can be written as the Laplace transform if $s = \lambda + i 2\pi\xi$:
\begin{align}
    \bar f(s) &\triangleq \mathcal L \left\{f(x)\right\} \triangleq  \int_0^\infty e^{-s x}  f(x)\;\mathrm  dx,
\end{align}
yielding the following relation between the Laplace and Fourier transform:
\begin{align}\label{eq:laplace_fourier}
    \mathcal L \left\{f(x)\right\} =  \mathcal F\left\{ e^{-\lambda x} f(x) u(x) \right\}.
\end{align}
If only positive values for $x$ are considered,
the inverse Laplace transform can be found via the inverse Fourier transform as well:
\begin{align}
    \mathcal L^{-1} \left\{\bar f(s)\right\}  &= f(x)\nonumber \\
    &=   e^{\lambda x} g(x; \lambda), \text{ from equation \eqref{eq:flt_gx}} \nonumber \\
    &= e^{\lambda x} \mathcal F^{-1} \left\{ \hat g(\xi; \lambda)  \right\} \nonumber \\
    &= e^{\lambda x} \mathcal F^{-1} \left\{ \bar f(\lambda + i2\pi\xi)  \right\},
\end{align}
therefore:
\begin{align} \label{eq:laplace_fourier_inv}
    f(x) = \mathcal L^{-1} \left\{\bar f(s)\right\} = e^{\lambda x} \mathcal F^{-1} \left\{ \bar f(\lambda + i2\pi\xi)  \right\}.
\end{align}

Next, the relations from equations \eqref{eq:laplace_fourier} and \eqref{eq:laplace_fourier_inv} are adapted to
the fast Fourier transform (FFT) to get the fast Laplace transform (FLT).
The fast Fourier transform algorithm is any algorithm that can compute the discrete Fourier transform and 
it's inverse in $\mathrm O(n\log n)$ complexity, the best know example is the algorithm presented by \cite{cooley1965algorithm}.
These FFT implementations are heavily optimized and available in many programming languages and frameworks.

To simplify the notation let $\{a\}_N = \{a_0, \dots, a_{N-1}\}$ be a sequence of numbers, and  
let $\alpha$, $\beta$ and $c$ be constants, the multiplication and addition operations on 
sequences are point-to-point operations defined as:
\begin{align*}
    \alpha\{a\}_N + \beta \{b\}_N &= \{\alpha a_0+\beta b_0,\dots, \alpha a_{N-1}+ \beta b_{N-1}\}, \\
    \{a\}_N + c &= \{a_0+ c,\dots, a_{N-1}+ c\}, \\
    \{a\}_N \{b\}_N &= \{a_0b_0,\dots,a_{N-1}b_{N-1}\}, \\
    \frac{\{a\}_N} {\{b\}_N} &= \left\{\frac {a_0} {b_0},\dots, \frac {a_{N-1}}{b_{N-1}}\right\}.
\end{align*}

Consider a sequence of $N$ samples of $f(x)$,
$\{f\}_N$, sampled on an equally spaced $\{x\}_N$ sequence such that 
$\rho = x_{i+1} - x_{i},\; \forall i \in \{0,\dots,N-2\}$.
Choose $\lambda$ big enough and create a new sequence
$\{g\}_N$, thus,
the discrete Laplace transform $\mathcal L_d \{f\}_N = \{\bar f\}_N$ of the input sequence $\{f\}_N$ is defined as the 
discrete Fourier transform $\{\hat g\}_N$
of the input $\{g\}_N$ as:
\begin{subequations}
    \label{eq:flt}
    \begin{align}
        g_n &= e^{-\lambda\frac{ n}{N}} f_n, \label{eq:gf_discrete}\\
        \bar f_k &= \hat g_k = \sum_{n=0}^{N-1} e^{-i2\pi k\frac{n}{N}} g_n,
    \end{align}
\end{subequations}
and for the inverse discrete Laplace transform  $\mathcal L^{-1}_d \{\bar f\}_N = \{f\}_N$  use the 
inverse discrete Fourier transform on $\{\bar f\}_N$ obtaining $\{g\}_N$
and invert the equation \eqref{eq:gf_discrete} to get $\{f\}_N$:
\begin{align}
    f_n &= e^{\lambda \frac{n}{N}}\frac{1}{N}\sum_{k=0} ^{N-1} e^{i2\pi k\frac{n}{N}} \bar f_k. \label{eq:iflt}
\end{align}
To work with these functions, those programming languages also provide functions to retrieve the discrete Fourier frequencies
$\xi_k$, with these frequencies, retrieve the Laplace frequencies $s_k$ so that:
\begin{subequations}
    \begin{align}
        \xi_k &= \frac{k}{N}, \\
        s_k &= \frac{\lambda}{N} + i2\pi \xi_k. \label{eq:flt_freq}
    \end{align}
\end{subequations}
Yielding the discrete Laplace transform and its inverse for a fixed $\lambda$:
\begin{subequations}
\begin{align}
    \bar f_k &= \sum_{n=0}^{N-1} e^{-s_kn} f_n,\\
    f_n &= \frac 1 N\sum_{k=0}^{N-1} e^{s_kn} \bar f_k.
\end{align}
\end{subequations}

\begin{thm} \label{thm:discrete_flt_derivative}
    For the same sampled space as above, the discrete Laplace transform of derivative of $f(x)$ is:
    \begin{align}
        \label{eq:discrete_flt_derivative}
        \mathcal L_d\{f'(x)\}_N =  \frac {\{s\}_N} \rho \mathcal L_d\{f'(x)\}_N - f_0.
    \end{align}        
\end{thm}

\subsection{Pricing call options}

The procedure stated next can be used to price European call options,
whether they are strike adjusted or not. It also can be used to price strike adjusted American calls
due to theorem \ref{thm:american_is_european}.
Special care should be taken for non-strike adjusted American calls once in this case we would have to
check for early exercise at each ex-dividend date.

To solve the series of PDEs in equation \eqref{eq:pde_recursive}, set the integration $x$ domain $[-L/2, L/2]$
to cover a significant amount of standard deviations $n_\sigma$
and subdivide this range into $N$ equally spaced points \footnote{$N$ is assumed to be even to simplify the discretization.}
which implies a sampling rate $\rho$, thus:
\begin{subequations}
    \begin{align}
        L &= 2n_\sigma \sigma \sqrt T, \\
        \rho &= \frac L N,\\
        x_n &= -\frac L 2 + n\rho,\; n \in \{0,\dots, N-1\}.
    \end{align}    
\end{subequations}
The sequence $\{x\}_N$ will be fixed for all time steps, on the other hand, the correspondent sequence $\{\tilde S_\tau\}$
will change obeying equation \eqref{eq:pde_change_Sx} accordingly:
\begin{align} \label{eq:s_tau_j_from_x}
        (\tilde S_\tau)_n = K_T\exp\left(x_n - \left(r - \frac{\sigma^2} 2\right)\tau\right),
\end{align}
with this, use the discretized version of equation \eqref{eq:call_ic} to get the sequence $\{F_1(x, 0)\}_N$ in $\tau = 0$:
\begin{align}
    F_1(x_n, 0) = K_T \max \{e^{x_n} -1, 0\}.
\end{align}
From here on apply the fast Laplace transform and its inverse which obey equations \eqref{eq:flt} and \eqref{eq:iflt}
respectively.

This transform needs um more parameter $\lambda$ which can be found by analyzing the convergence of
$e^{-\lambda x} f(x) \to f(x_0)$ when $x \to \infty$, which gives the following approximation:
\begin{align}
    e^{-\lambda \frac{N-1} N} F_1(x_{N-1}, 0) = F_1(x_0, 0)
    \implies \lambda =\frac N {N-1} \ln\left(\frac {F_1(x_{N-1}, 0)}{F_1(x_0, 0)}\right).
\end{align}

Next, follow as described in section \ref{sec:pricing_strategy} applying the discretized
versions of equations \eqref{eq:pde_s_solution} to \eqref{eq:f_dividend_remapping}.
\begin{subequations}
    \begin{align}
        \bar F_i(s_k, \tau_i)
        &=  \exp\left( \frac{(\sigma s_k)^2} {2\rho^2}  (\tau_i - \tau_{i-1}) \right)
        \bar F_i(s_k, \tau_{i-1}^-),   \label{eq:pde_s_solution_discrete} \\
        F_i(\{x\}_N, \tau_i)
        &=  \mathcal L_d^{-1}\left\{ \bar F_i(\{s\}_N, \tau_i) \right\},\label{eq:pde_x_solution_discrete}
    \end{align}
\end{subequations}
and to do the remapping from equation \eqref{eq:f_dividend_remapping}
interpolate the sequence $\{x\}_N$ using the mapping $(x_{\tau^-})_j \to F_i(x_j, \tau_i)$
to find the sequence $F_{i-1}(\{x\}_N,{\tau_i^-})$, where $(x_{\tau^-})_j$ can be
described as accounting for the dividend change in
$\tilde S_\tau$ from equation \eqref{eq:s_tau_j_from_x}, thus:
\begin{align}
    (x_{\tau^-})_j = \ln\left(\frac{(\tilde S_\tau)_j + D_{n+1 -j}}{K_T}  \right) + \left(r - \frac{\sigma^2} 2\right)\tau.
\end{align}
Iterate, solving each $F_i(\{x\}_N,\tau)$ until the final mapping $\{x\}_N \to F_{n+1}(\{x\}_N,T)$ is reached,
also find the first and second derivatives with respect to $x$ via equations \eqref{eq:fx_derivatives}, namely:
$\partial_x F_{n+1}(\{x\}_N, T)$ , and $\partial_{xx} F_{n+1}(\{x\}_N, T)$,
in which $x_T$  can be interpolated to find $F_{n+1}(x_T,T)$, $\partial_x F_{n+1}(x_T, T)$, and $\partial_{xx} F_{n+1}(x_T, T)$, so that:
\begin{align}
    C_t &= F_{n+1}(x_T,T) \exp(-rT),\\
    \Delta_t &= \frac 1 S_t \frac{\partial F_{n+1}(x_T, T)} {\partial x} \exp(-rT),\\
    \Gamma_t &= \frac 1 {S_t^2} \left(  \frac{\partial^2 F_{n+1}(x_T, T)} {\partial x^2} -  \frac{\partial F_{n+1}(x_T, T)} {\partial x}\right)\exp(-rT),\\
    \Theta_t &= C_t r - S_t r\Delta_t - \frac {\sigma^2 S^2} 2 \Gamma_t,
\end{align}
will give the vanilla call price and Greeks.

\section{Results}
\label{sec:results}

\begin{figure}[H]
    \centering
    \includegraphics[width=\textwidth]{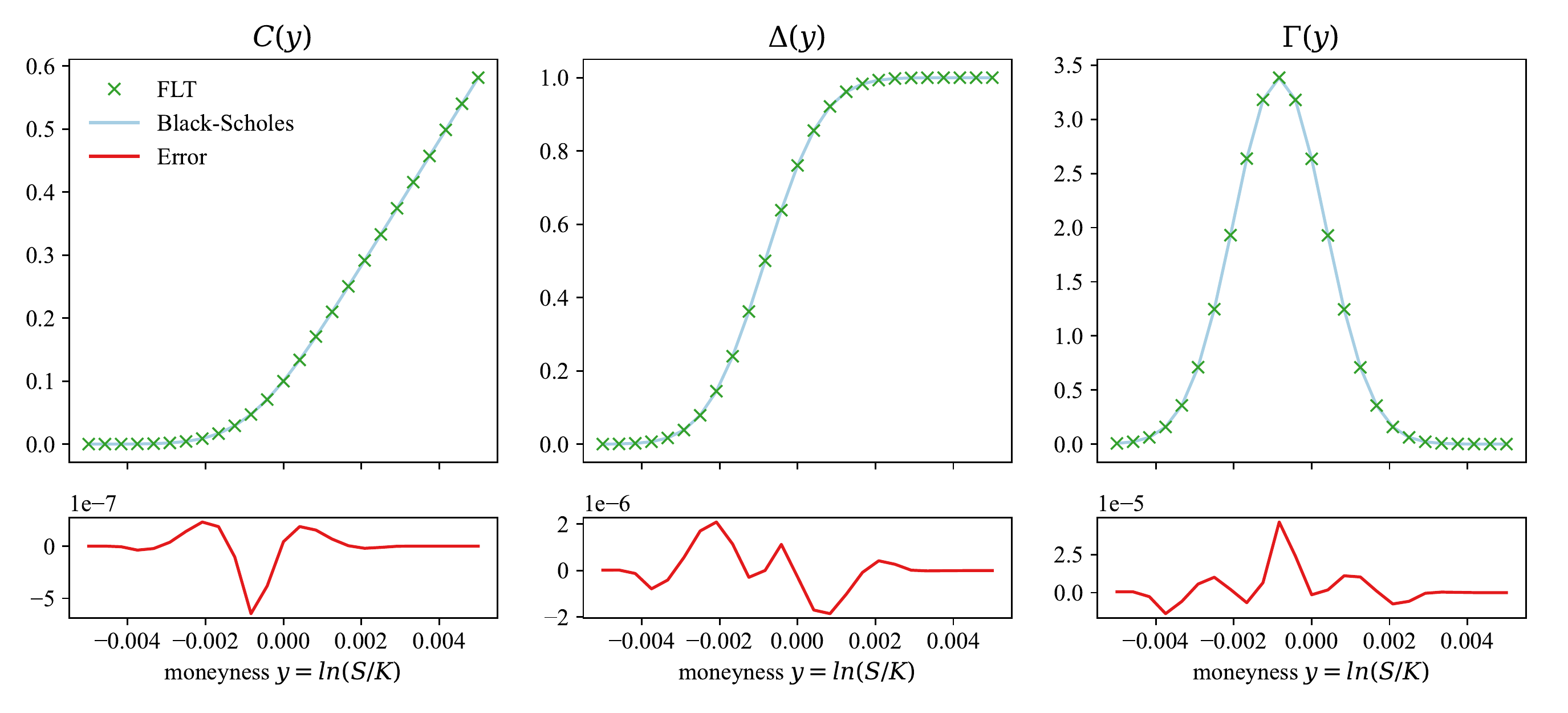}
    \caption{Premium $C(y)$, and Greeks $\Delta(y)$ and $\Gamma(y)$,
            $S_0 = 100$, $r = 6\%$ , $\sigma = 1\%$, and $T = 5$ days.}
    \label{fig:bs_baseline}
\end{figure}

To establish a baseline\footnote{The source code used to produce the following results can be found at \url{https://github.com/maikonaraujo/paper_202209}.}, 
the first step is to test the model in the condition where there is no dividend
being paid up to the expiry. In this condition, the model should deliver the same price and Greeks as
the plain vanilla call formulas from the Black-Scholes model. Figure \ref{fig:bs_baseline} compares the
premium and Greeks for a range of moneyness using the fast Laplace transform pricer in green against
the Black-Scholes results in blue, showing the error below in red. To stress the model the volatility was set to a very low
level of $\sigma = 1\%$ in a very close to maturity option ($T = 5$ days), and the Laplace transform
was discretized with $N=2^{10}$ equally spaced points. In this section,
all FLT pricers use $n_\sigma = 7.5$. 

For the case where only one dividend $D$ is paid  is paid at time $t$ for an option expiring in
$T=1$ year and the market conditions are $S_0=100$, $r=6\%$, $\sigma=30\%$, table \ref{tbl:thakoor_bhuruth}
compares the values of the FLT pricer with the \cite{THAKOOR20181} European pricer (TB)
for dividends paid very close to the reference date ($t=0.0001$), $6$ months ($t=0.5$)
and very close to expiry ($t=0.9999$) for in-the-money, at-the-money, and  out-off-the-money strikes
$K=70$, $K=100$, and $K=130$ respectively. The prices from the FLT pricer agree with the ones from TB pricer.
The FLT was discretized with $N=1024$ equally spaced points, and TB uses $N=500$ integration points with $\xi=6.5$
standard deviations. Both prices have the final strike adjusted to account for the dividend payment.

\begin{table}[H]
    \centering
    \caption{One dividend payment - Comparison with Thakoor, and Bhuruth model.}
    \label{tbl:thakoor_bhuruth}
    \begin{footnotesize}
        \begin{tabular}{@{}ccSSSSSSSSS@{}}
            \toprule
                  &  &\multicolumn{3}{c}{$K=70$}&\multicolumn{3}{c}{$K=100$}&\multicolumn{3}{c}{$K=130$}
            \\\cmidrule(lr){3-5}\cmidrule(lr){6-8}\cmidrule(lr){9-11}
            t     &D &FLT                       &TB                         &{Diff.}                    &FLT    &TB     &{Diff.}    &FLT   &TB    &{Diff.} \\\midrule
            0.0001&7 &34.3193                   &34.3193                    &{2.76e-12}                 &13.6870&13.6870&{9.63e-13} &4.1862&4.1862&{3.19e-13} \\
            0.5000&7 &34.6419                   &34.6419                    &{3.55 e-14}                &14.2172&14.2172&{1.42e-13 }&4.5808&4.5808&{3.11e-14 } \\
            0.9999&7 &34.9844                   &34.9844                    &{2.55 e-08}                &14.7170&14.7170&{2.39e-08 }&4.9195&4.9195&{2.32e-08 } \\\midrule
            0.0001&20&33.1952                   &33.1952                    &{4.76 e-13}                &11.7740&11.7740&{3.45e-13 }&2.9221&2.9221&{3.46e-14 } \\
            0.5000&20&34.0504                   &34.0504                    &{3.55 e-14}                &13.3326&13.3326&{1.49e-13 }&4.0013&4.0013&{2.13e-14 } \\
            0.9999&20&34.9842                   &34.9842                    &{2.55 e-08}                &14.7168&14.7168&{2.44e-08 }&4.9194&4.9194&{2.32e-08 } \\\midrule
            0.0001&50&31.1664                   &31.1664                    &{2.38 e-13}                &7.3596 &7.3596 &{1.10e-13 }&0.7210&0.7210&{2.44e-15 } \\
            0.5000&50&32.9077                   &32.9077                    &{1.84e-06}                 &11.5707&11.5707&{5.51e-14 }&2.9205&2.9205&{4.13e-14 } \\
            0.9999&50&34.9840                   &34.9840                    &{2.55e-08 }                &14.7165&14.7165&{2.44e-08 }&4.9192&4.9192&{2.32e-08 } \\\bottomrule
        \end{tabular}
    \end{footnotesize}
\end{table}

Starting from the same market conditions and strikes used in table \eqref{tbl:thakoor_bhuruth}, 
table \eqref{tbl:option_greeks} adds two cases with multiple dividends and compares the Greeks 
from the FLT pricer, namely $\Delta$, and $\Gamma$, against Greeks from a numeric bump in the stock price of $h=0.01$, so that:
\begin{align}
    \Delta_n  &= \frac{\mathrm{FLT}(S + h) - \mathrm{FLT}(S - h)}{2h}, \\
    \Gamma_n  &= \frac{\Delta(S + h) - \Delta(S - h)}{2h}.
\end{align}
For these simulations, the FLT uses $N=100$ equally spaced points. 
It can be seen that the numerical Greeks agree with the ones from the numeric bump central derivative.

\begin{table}[H]
    \centering
    \caption{Greeks - Analytical and numerical bumps.}
    \label{tbl:option_greeks}
    \begin{footnotesize}
        \begin{tabular}{@{}cSSSSSSSSSS@{}}
            \toprule
               &\multicolumn{5}{c}{$(t_i,D_i) = \{(0.2, 4), (0.4,5) , (0.6, 6), (0.8, 3)  \}$ }&\multicolumn{5}{c}{$(t_i,D_i) = \{(0.2, 9), (0.6,9) \}$ }
            \\ \cmidrule(lr){2-6}\cmidrule(lr){7-11}
            K  &FLT &{$10^2\Delta$}   &{$10^2\Delta_{n}$}&{$10^4\Gamma$}&{$10^4\Gamma_n$}& FLT  &{$10^2\Delta$}&{$10^2\Delta_{n}$}&{$10^4\Gamma$}&{$10^4\Gamma_n$} \\ \midrule
            70 & 34.1131 & 95.41 & 95.41 & 35.99 & 35.99  & 33.9703 & 95.69 & 95.69 & 34.83 & 34.83 \\
            100 & 13.4083 & 63.41 & 63.41 & 137.87 & 137.87  & 13.1728 & 63.41 & 63.41 & 140.34 & 140.34 \\
            130 & 4.0395 & 27.40 & 27.40 & 120.74 & 120.74 & 3.8780 & 26.91 & 26.91 & 121.80 & 121.80 \\
            \bottomrule
        \end{tabular}
    \end{footnotesize}
\end{table}

\begin{figure}[H]
    \centering
    \includegraphics[width=\textwidth]{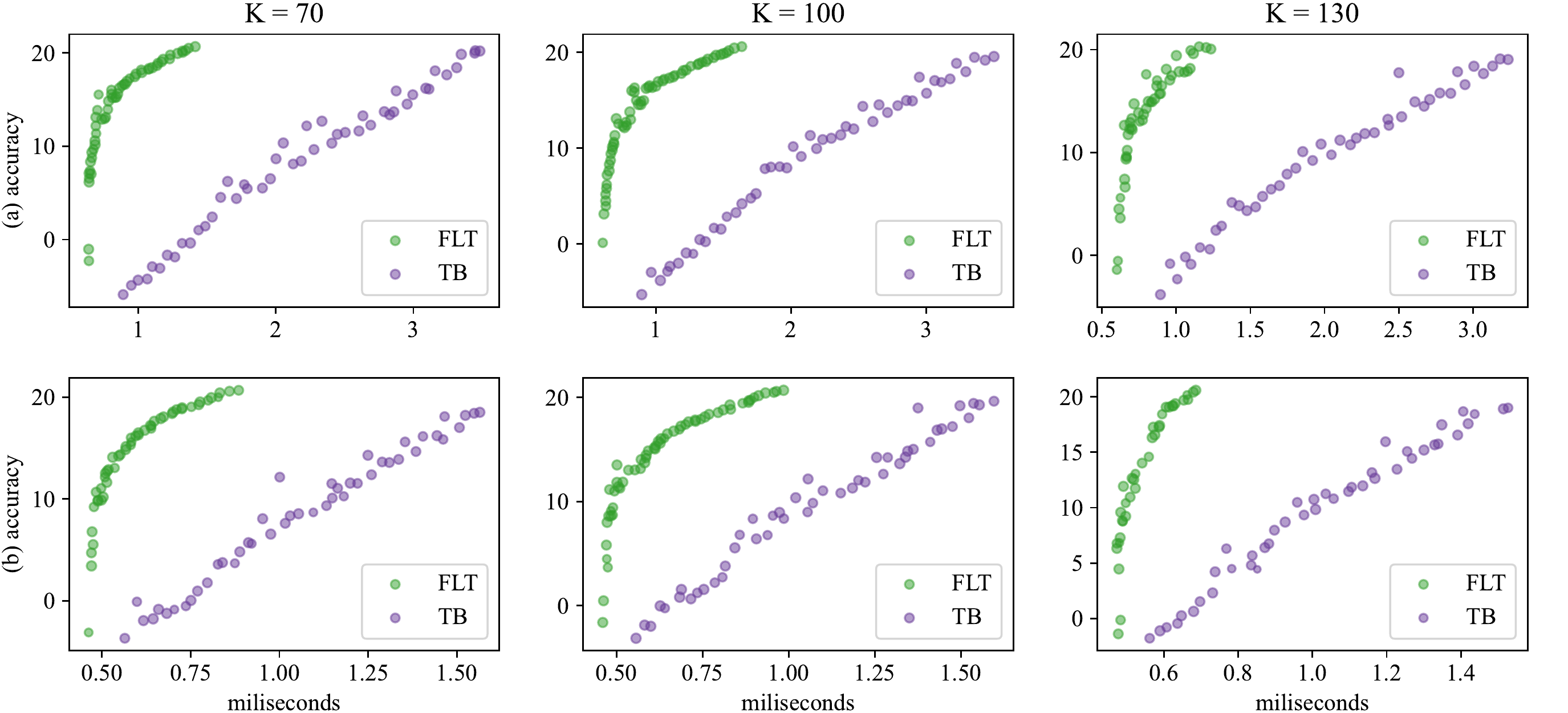}
    \caption{Time and Accuracy comparison. 
        (a)  $(t_i,D_i) = \{(0.2, 4), (0.4,5) , (0.6, 6), (0.8, 3)  \}$. 
        (b) $(t_i,D_i) = \{(0.2, 9), (0.6,9) \}$.}
    \label{fig:time_accuray}
\end{figure}

Concerning performance, when compared to the TB pricer, the FLT pricer presents better results because it reaches the same level of accuracy faster \footnote{Both models were implemented in python, a highly optimized C/C++ implementation could modify this result.}
and has the advantage of also delivering the Greeks $\Delta$, $\Gamma$, and $\Theta$ together with the option's premium. 
Figure \ref{fig:time_accuray} shows some simulations for time versus accuracy.
The line (a) shows the four dividend case $(t_i,D_i) = \{(0.2, 4), (0.4,5) , (0.6, 6), (0.8, 3)  \}$ and the line (b) shows the two dividend case 
$(t_i,D_i) = \{(0.2, 9), (0.6,9) \}$. The columns show strikes $70, 100,$ and $130$. To estimate a true value both prices were simulated with extreme discretizations 
finding prices $\hat p_a$ and $\hat p_b$ which agree with each other with an error of order (1.0E-11) for the case (a) and order (1.0E-14) for the case (b). 
The accuracy is measured by:
\begin{align}
    \text{Accuracy}_i &= - \log\left(|p_i - \hat p|\right), \\    
\end{align}
for each simulation. The time taken is measured by running 10 batches of 25 evaluations,
and selecting the best average time, therefore, if the batch took time $b_j$ for $(j \in 1,\dots,10)$, $t_i$ is:
\begin{align}
    t_i = \frac{\min\{b_j\}}{25}.
\end{align}
To give a sense of variance of the time measured, the size $s_i$ of each marker in figure \ref{fig:time_accuray} is proportional to:
\begin{align}
    s_i = \frac{\min\{b_j\}}{\max\{b_j\}}
\end{align}
thus simulations with greater variance have smaller markers. Everything was run on a PC with 16GB of RAM, an intel i7 processor with 2.60GHz, 6 cores, 
the L1, L2, and L3 cache sizes are 384KB, 1.5MB, and 12MB respectively.

\section{Conclusion}

This work has presented a numerical procedure able to price and hedge options with 
discrete multiple dividends using the fast Laplace transform. The FLT is a great tool for solving 
differential equations and delivers great performance and stability when applied to options pricing 
with constant volatility. Brazilian listed equity options represent a significant¢ market and are 
a perfect fit for this approach, but other markets can also benefit from this methodology.

\appendix
\section{Proofs of theorems and corollaries}

\begin{pf}[Theorem \ref{thm:black_scholes_pde}]
    Let $\pi_t$ be a self-financing portfolio of a long position in a security $V_t$, a short position 
    on the stock price $S_t$, and a cash account to invest the received dividends up to time $t$: $I_t$. 
    The value of this portfolio at any time is given by:
    \begin{align}
        \pi_t  = V_t - \Delta_t (S_t + I_t).
    \end{align}
    The derivative of this portfolio is given by:
    \begin{align}
        \mathrm d\pi_t  &= \frac{\partial V_t}{\partial t} \mathrm  dt + \frac{\partial V_t}{\partial S_t}\mathrm  dS_t + \frac{\mathrm d\left\langle S_t,S_t\right\rangle }{2}  \frac{\partial^2 V_t}{\partial S_t^2} 
        + \mathcal J V_t
        - \Delta_t (\mathrm dS_t +\mathrm  dI_t)
    \end{align}
    Assuming that the security value is continuous for all $t\in [0,T]$, we have $\mathcal J V_t = V_t(S_t) - V_{t^-}(S_{t^-}) = 0, \,\forall t \in [0, T]$,
    additionally with equations \eqref{eq:s_process_with_dividend} and \eqref{eq:cash_past_dividends} follows that:
    \begin{align}
        \mathrm d\pi_t  &= \frac{\partial V_t}{\partial t}\mathrm   dt 
        + \frac{\partial V_t}{\partial S_t} (S_t r_t\mathrm  dt + S_t\sigma_t\mathrm  dW_t) 
        + \frac{(\sigma_t S_t)^2 }{2}  \frac{\partial^2 V_t}{\partial S_t^2} \mathrm dt
        - \Delta_t \left((S_t+I_t) r_t\mathrm  dt + S_t \sigma_t\mathrm  dW_t \right),    
    \end{align}
    choosing $\Delta_t = \frac{\partial V_t}{\partial S_t}$ we hedge the risk $dW_t$ away, turning  $\pi_t$ into a risk free 
    portfolio hence $d\pi_t = \pi_t r_t dt$. Moreover:
    \begin{align}
        \pi_t r_t \mathrm dt  &= \frac{\partial V_t}{\partial t}\mathrm   dt     
        + \frac{(\sigma_t S_t)^2}2  \frac{\partial^2 V_t}{\partial S_t^2} \mathrm  dt
        - \Delta_t I_t r_t\mathrm  dt ,    \\
        (V_t -\Delta_t(S_t + I_t)) r_t  &= \frac{\partial V_t}{\partial t}  
        + \frac{(\sigma_t S_t)^2}2  \frac{\partial^2 V_t}{\partial S_t^2} 
        - \Delta_t I_t r_t ,    \\        
        0 &= \frac{\partial V_t}{\partial t}  
            + \frac{(\sigma_t S_t)^2}2  \frac{\partial^2 V_t}{\partial S_t^2} 
            +  S_t r_t \frac{\partial V_t}{\partial S_t} -  r_tV_t.
        \end{align}
        \qed
    \end{pf}

    \begin{pf}[Corollary \ref{coro:option_martingale}]
        Taking the derivative of $X_t = e^{-\int_0^t r_u\;\mathrm  du}V_t$ we get:
        \begin{align}
            \mathrm dX_t & = -r_t e^{-\int_0^t r_u\;\mathrm  du}V_t \mathrm  dt +  e^{-\int_0^t r_u\;\mathrm  du}\mathrm  dV_t \nonumber \\
        &= e^{-\int_0^t r_u\;\mathrm  du} \left( -rV_t \mathrm  dt 
            + \frac{\partial V_t}{ \partial t} \mathrm dt 
            + \frac{1}{2}(\sigma_tS_t)^2\frac{\partial^2 V_t}{\partial S_t^2}\mathrm dt 
            + \frac{\partial V_t}{\partial S_t}(S_tr_t \mathrm dt + S_t\sigma_t \mathrm dW_t)
               \right),
        \end{align}
        from theorem \ref{thm:black_scholes_pde} we know that the terms multiplying $dt$ add to zero, thus:
        \begin{align}
            \mathrm dX_t & = e^{-\int_0^t r_u\; \mathrm du}\frac{\partial V_t}{\partial S_t}S_t\sigma_t \mathrm  dW_t \nonumber  \\
                    & = S_t\sigma_t\frac{\partial X_t}{\partial S_t}\mathrm  dW_t,
        \end{align}
        proving that $X_t$ is a martingale. \qed
    \end{pf}

    \begin{pf}[Theorem \ref{thm:american_is_european}]
        At any time $t$ the option holder can decide between exercising the option, hence receiving its intrinsic 
        value $g_i(S_t)$ or keeping the option. Therefore, the value of the American option at any 
        time $t$ can be described as:
        \begin{align}
            C_{t} = \max \left\{  \mathbb E\left[ \left. e^{-\int_t^u r_v\;\mathrm  dv}  C_{u>t}  \right| \mathcal F_t\right] , g_n(S_t)\right\} ,
        \end{align}
        where $t < u <= T$. To prove that it is never optimal to exercise, it suffices to prove that for each $t \in [t_i, t_{i+1})$:
        \begin{align}
        \mathbb E\left[ \left. e^{-\int_t^{t_{i+1}} r_v\; \mathrm dv}  C_{t_{i+1}}  \right| \mathcal F_t\right] \ge g_i(S_t).
        \end{align}
        Let's prove it by backward induction. Starting with the last period where $t \in [t_n, T]$, we have:
        \begin{align}
            \mathbb E\left[ \left. e^{-\int_t^T r_v\;\mathrm dv}  g_n(S_T)  \right| \mathcal F_t\right] 
            &\ge \mathbb E\left[ \left.  g_n\left(e^{-\int_t^T r_v\;\mathrm  dv} S_T\right)  \right| \mathcal F_t\right]  \text{ (lemma \ref{lemma:call_payof_convex}),}\nonumber \\
            &\ge g_n\left(\mathbb E\left[ \left.  e^{-\int_t^T r_v\;\mathrm  dv} S_T  \right| \mathcal F_t\right]\right) \text{ (Jensen's inequality),}\nonumber \\
            &= g_n\left(S_t\right) \text{ (martingale property).} \label{eq:last_dividend_call_intrinsic}
        \end{align}
        Now, let's take an instantaneous step back in time for the exact moment when the last dividend goes ex-date, then:
        \begin{align}
            C_{t_n^-} &= max \left\{ C_{t_n} , g_{n-1}(S_{t_n^-})  \right\} \nonumber \\
                    &=  max \left\{ C_{t_n} , g_n(S_{t_n})  \right\}  \text{ (equation \eqref{eq:call_payof_continuous})}, \nonumber \\
        \therefore \, C_{t_n^-} &= C_{t_n} \text{ (by equation \eqref{eq:last_dividend_call_intrinsic})}.
        \end{align}
        To complete the inductive argument, consider a time $t \in [t_k, t_{k+1})$, our inductive hypothesis is:
        \begin{align}
            C_{t_{k+1}^-}  &\ge  g_k(S_{t_{k+1}^-}), 
        \end{align}
        from here we can proceed in the same manner as we did deriving equation \eqref{eq:last_dividend_call_intrinsic}:
        \begin{align}            
            \mathbb E\left[\left. e^{-\int_t^{t_{k+1}} r_u\;\mathrm du} C_{t_{k+1}^-} \right| \mathcal F_t\right]   &\ge  \mathbb E\left[ \left.e^{-\int_t^{t_{k+1}} r_u\;\mathrm  du} g_k(S_{t_{k+1}^-}) \right| \mathcal F_t\right] \nonumber \\
                &\ge  \mathbb E\left[ \left.  g_k\left(e^{-\int_t^{t_{k+1}} r_u \;\mathrm  du}S_{t_{k+1}^-}\right) \right| \mathcal F_t\right] \nonumber \\ 
                &\ge  g_k \left(\mathbb E\left[ \left.  e^{-\int_t^{t_{k+1}} r_u\;\mathrm  du}S_{t_{k+1}^-} \right| \mathcal F_t\right] \right)\nonumber \\ 
                &=  g_k \left( S_t \right),
        \end{align}
        which proves the induction. \qed        
        \begin{lemma} \label{lemma:call_payof_convex}
            For any $g(x) = \max \left\{ x - K, 0\right\}$, $x,K > 0$ we have:
            \begin{align}
                g(a x) \le ag(x), 
            \end{align}
            when $a \in [0, 1]$. The proof is a direct application of Jensen's inequality to the convex function 
            $g(x)$.
        \end{lemma}
    \end{pf}

    \begin{pf}[Theorem \ref{thm:forward_price}] 
        Fist, note that the process $X_t = e^{-\int_0^{t} r_u\;\mathrm  du}(S_t + I_t)$ is a martingale:
        \begin{align}
            \mathrm dX_t &= -r_t X_t \mathrm dt + e^{-\int_0^{t} r_u\; \mathrm du}(\mathrm dS_t + \mathrm dI_t) \nonumber \\
                 &= -r_t X_t\mathrm  dt + e^{-\int_0^{t} r_u\;\mathrm  du}(r_tS_t \mathrm dt + \sigma_t S_t\mathrm  dW_t + \mathcal J S_t + rI_t \mathrm dt +\mathcal J I_t),  \nonumber \\
                 &= -r_t X_t \mathrm  dt + e^{-\int_0^{t} r_u\;\mathrm  du}(r_tS_t \mathrm dt + \sigma_t S_t\mathrm  dW_t + rI_t\mathrm  dt) \text{, because }(\mathcal J S_t = - \mathcal J I_t),\nonumber \\                 
                 &= -r_t X_t\mathrm  dt + + r_tX_t \mathrm dt + e^{-\int_0^{t} r_u\;\mathrm  du}\sigma_t S_t\mathrm  dW_t , \nonumber \\
                 &= e^{-\int_0^{t} r_u\;\mathrm  du}\sigma_t S_t\mathrm  dW_t.
        \end{align}            
        Therefore:
        \begin{align}
            \mathbb E\left[\left. e^{-\int_0^T r_u\;\mathrm  du}(S_T + I_T) \right| \mathcal F_t \right] = e^{-\int_0^t r_u\;\mathrm  du}(S_t + I_t) ,\nonumber \\
            \mathbb E\left[\left. S_T  \right| \mathcal F_t \right] = e^{\int_t^T r_u\;\mathrm  du}(S_t + I_t) - I_T.
        \end{align}               
        \qed
    \end{pf}

    \begin{pf}[Theorem \ref{thm:discrete_flt_derivative}] 
        Letting $x(n)= \rho n + x_0$ and taking the continuous derivative of $f(x)u(x-x_0)$ with respect to $x$:
        \begin{align}
            [f(x)u(x - x_0)]' &= f'(x)u(x-x_0) + f(x)\delta(x-x_0) \implies \\
             \frac {\mathrm d [f(n)u(n)]}{\mathrm d n} \frac {\mathrm d n}{\mathrm d x} &=  \frac {\mathrm df(x)} {\mathrm dx} u(n) + f(n) \mathbf 1_{n=0},
        \end{align}
        then applying the discrete Laplace transform to both sides:        
        \begin{align}\label{eq:proof_lft_deriv_eq1}            
                \sum_{n=0}^{N-1} e^{-s_kn} \frac {\mathrm d [f(n)u(n)]}{\mathrm d n} \frac {\mathrm d n}{\mathrm d x} 
                &=\sum_{n=0}^{N-1} e^{-s_kn} \frac {\mathrm df(x)} {\mathrm dx} u(n) + \sum_{n=0}^{N-1} e^{-s_kn}  f(n) \mathbf 1_{n=0}, \nonumber \\
                &=\sum_{n=0}^{N-1} e^{-s_kn} \frac {\mathrm df(x)} {\mathrm dx} + f_0,
        \end{align}
        
        The left-hand side of equation \eqref{eq:proof_lft_deriv_eq1} can be written with the inverse Laplace transform:
        \begin{align}            
            \sum_{n=0}^{N-1} e^{-s_kn} \frac {\mathrm d [f(n)u(n)]}{\mathrm d n} \frac {\mathrm d n}{\mathrm d x} 
               &= \sum_{n=0}^{N-1} e^{-s_kn}  \frac {\mathrm d\left( \frac 1 N \sum_{l=0}^{N-1} e^{s_l n} \bar f_l \right)} {\mathrm dn}\frac {1}{\rho}   
               = \sum_{n=0}^{N-1} e^{-s_kn}  \left( \frac 1 N \sum_{l=0}^{N-1} e^{s_l n} s_l\bar f_l \right)\frac {1}{\rho}  \nonumber \\
               &= \frac 1 {N\rho} \sum_{n=0}^{N-1}\sum_{l=0}^{N-1} e^{(s_l-s_k) n} s_l\bar f_l 
               = \frac 1 {N\rho} \sum_{l=0}^{N-1} s_l\bar f_l\sum_{n=0}^{N-1} e^{\frac{i2\pi(l-k) n}{N}} \nonumber \\
               &= \frac 1 {N\rho} \sum_{l=0}^{N-1} s_l\bar f_l N\mathbf 1_{l=k} 
               = \frac {s_k} {\rho} \bar f_k.
       \end{align}
       Taking sequences from $k \in \{0, \dots, N-1\}$ and rewriting equation \eqref{eq:proof_lft_deriv_eq1} in terms of the 
       discrete Laplace transform finishes the proof:
       \begin{align}
            \frac {\{s\}_N} {\rho} \mathcal L_d\{f\}_N &= \mathcal L_d\{f'(x)\}_N + f_0, \nonumber\\
            \therefore \mathcal L_d\{f'(x)\}_N &=\frac {\{s\}_N} {\rho} \mathcal L_d\{f\}_N - f_0.
       \end{align}
       \qed
    \end{pf}




\bibliographystyle{elsarticle-harv} 
\bibliography{refs}

\end{document}